\documentclass[aps,preprint,groupedaddress]{revtex4}
\usepackage{amsmath, amsthm, amssymb}

\newcommand{\mnras}{{\it Monthly Not.\ Roy.\ Astron.\ Soc.}}

\newtheorem{thm}{Theorem}
\newtheorem{cor}{Corollary}

\begin{document}

\draft
\title{Timelike Geodesic Currents in the Stationary, Axisymmetric,  Force-free Magnetosphere of a Kerr Black Hole}
\author{Govind Menon}
\address{Dept.\ of Mathematics and Physics,\\
Troy University, Troy, Alabama, 36082}
\author{Charles D.\ Dermer}
\address{E. O. Hulburt Center for Space Research, Code 7653,
Naval Research Laboratory, Washington, DC 20375-5352}
\date{\today}
\begin{abstract}
The structural properties of geodesic currents in an ambient Kerr background is studied from an analytical point of view. The geodesics in the congruence correspond to charged particles that carry energy and angular momentum from the black hole through the Blandford-Znajek mechanism. It is shown that the resulting magnetosphere naturally satisfies the Znajek regularity condition. Particular attention is paid here to the energy extracted by matter currents rather than by electromagnetic Poynting fluxes.
\end{abstract}


\maketitle Force-free Electrodynamics of Kerr Black holes
\section{Introduction}
Nearly thirty years ago, Blandford and Znajek \cite{bz77} described a method by which a black hole could release its 
rotational energy into the magnetosphere via electromagnetic processes. They argued for the existence of a force-free magnetosphere, wherein outflowing currents and electromagnetic energy fluxes could in principle carry energy and angular momentum away from the black hole.
Subsequently, a considerable amount of research has been done in trying to understand the nature of the solutions provided by a force-free electrodynamics in the vicinity of a rotating black hole (e.g., \cite{mdt82,tmd82,tpm86,pun01,kom04}). Such solutions are extremely important from an astrophysical point of view, as they would be leading candidates for the explanation of jets emanating from black holes. Owing to the complexity of the problem, recent efforts have been primarily numerical in nature (e.g., \cite{sdp04,mn07}).

After obtaining an approximate solution to the Blanford-Znajek mechanism  of energy extraction from rotating black holes that generalizes the original perturbative split monopole solution \cite{md05},  we  were recently successful \cite{md07}
in providing a class of exact analytic solutions to the equations of force-free electrodynamics in Kerr geometry. Although finite everywhere in the magnetosphere, these solutions were not, however, physically realistic, but provide a basis for further analytic study.
For the case of stationary, axisymmetric, force-free electrodynamics in a Kerr background, 
we \cite{md07} showed that  that the exact solution permitted an electromagnetic current 4-vector of the form
\begin{equation}
I^{\nu} =-\frac{2}{a^2  \sqrt{-g}} \frac{d}{d \theta}[\Lambda \frac{\cos{\theta}}{\sin^4{\theta}}]\;l^{\nu}\;.
\label{oldcurrent}
\end{equation}
Here, $a $ is the angular momentum per unit mass of the black hole, 
$g$ is the usual determinant of the Kerr spacetime metric $g_{\mu \nu}$,
and $\Lambda$ is an arbitrary function of $\theta$. Also $l^\nu$ is the infalling principle null geodesics of the Kerr geometry, given explicitly in Boyer-Lindquist coordinates $\{t,r,\theta,\varphi\}$ by 
\begin{equation}
l^{\nu} = (\frac{r^2+a^2}{\Delta}, -1, 0, \frac{a}{\Delta}).
\label{princenullgeo}
\end{equation}
Here, $\Delta = r^2+a^2-2Mr$, where $M$ is the mass of the black hole. Since the geodesic in eq.\ (\ref{princenullgeo}) is infalling, it is not possible to extract energy from the hole. More importantly, the current flows along null geodesics.  While we expect the currents to flow through geodesics under force-free conditions, the fact that the geodesics are null is a less desirable feature.

In this paper, we study the general structural properties of geodesic currents of particles with mass when the magnetosphere is force-free. 
Instead of working with the covariant formalism, we will write all
equations of electrodynamics in an intuitive $3+1$ formalism
\cite{kom04}. After a brief introduction to the $3+1$ formalism in
Section \ref{3+1}, we describe the relevant equations of a
force-free, axisymmetric, stationary magnetosphere of a Kerr black
hole in Section \ref{fofee}. The structural properties of geodesic
currents are described in Section \ref{geocurrent}, where we demonstrate that the 
Znajek regularity condition is satisfied. Energy and angular momentum extraction 
through particle currents in the force-free magnetosphere are considered in
Section \ref{particlejets}, and we show in 
Section  \ref{sectarea} that a naked singularity cannot result from this process.
We conclude in Section \ref{conclusions}. 


\section {3+1 Electrodynamics in Kerr Geometry: a primer}
\label{3+1} In the 3+1 formalism of electrodynamics, we rewrite the covariant Maxwell equations in terms of 3-vectors $E$, $B$, $D$, and $H$. While these quantities may loose meaning under arbitrary coordinate transformations, in a particular frame we have the advantage of working with the four familiar equations of electrodynamics. 
Much of the background and details presented in this section can be found in \cite{kom04} and \cite{md05}.

The 3-vectors are defined on spacelike slices of spacetime. We shall
foliate the Kerr spacetime by a collection of absolute space
described by surfaces of constant time $t$ in the Boyer-Lindquist
coordinates $\{t,r,\theta,\varphi\}$ of the Kerr geometry. Here, the metric takes the form
\begin{equation}
ds^2=( \beta^2 - \alpha^2 )dt^2 \;+  \;2 \;\beta_\varphi d\varphi
dt\;+\gamma_{rr} dr^2 + \;\gamma_{\theta \theta} d\theta^2 +
\;\gamma_{\varphi\varphi}d\varphi^2 . \label{ds2}
\end{equation}
The metric coefficients are given by 
$$\beta^2-\alpha^2 \;= \;g_{tt} \;=\; -1 + \frac{2Mr}{\rho^2}\;,\;\;\;\beta_\varphi \; \equiv g_{t \varphi}\; = \;\frac{-2Mr a
\sin^2\theta}{\rho^2}\;,\;\;\;\gamma_{rr} =
\frac{\rho^2}{\Delta}\;,$$
\begin{equation}
\gamma_{\theta \theta} = \rho^2, \;\; {\rm and}\;\; \gamma_{\varphi \varphi} = \frac{\Sigma^2 \sin^2\theta}{\rho^2}\;,
\label{kerrterms}
\end{equation}
where \begin{equation} \rho^2 = r^2 + a^2
\cos^2\theta\;,\;\;\;\Delta = r^2 -2 M r + a^2\;\;\;{\rm and}\;\;\;
\Sigma^2 = (r^2 + a^2)^2 -\Delta \; a^2 \sin^2\theta\;.
\end{equation}
 Additionally
\begin{equation}
\alpha^2 = \frac{\rho^2 \Delta}{\Sigma^2}, \;\;\; \beta^2 =
\frac{\beta_\varphi^2}{\gamma_{\varphi \varphi}}\;,\;{\rm
and}\;\;\sqrt{-g}=\alpha\; \sqrt{\gamma} = \rho^2 \sin\theta\;.
\label{metfcnalpha}
\end{equation}
Parameters $M$ and $a$ are the mass and angular momentum per unit
mass respectively of the Kerr black hole. The three-dimensional
space for some fixed value of $t$ is endowed with the metric
\begin{equation}
ds_{sp}^2= \gamma_{rr} dr^2 + \gamma_{\theta\theta} d\theta^2
+ \gamma_{\varphi\varphi}d\varphi^2.
\label{spacemet}
\end{equation}
Since the spacetime is
stationary, the value of $\;t$ for the absolute space will not matter
in our discussion.
The covariant Maxwell's equations are
\begin{equation}
\nabla_\beta ^\star F^{\alpha \beta} = 0 \;, \;{\rm and} \;
\nabla_\beta  F^{\alpha \beta} = I^\alpha\;.
\end{equation}
Here $F^{\alpha \beta}$ is the Maxwell stress tensor, and $I^\alpha$ is the four vector of the electric current. In order to rewrite the covariant equations of electrodynamics in a familiar Maxwell-type form, three vectors $D$ and $H$ in our absolute space are defined by
\begin{equation}
D^i=\alpha  F^{ti} \;, \;{\rm and} \; H_i = \frac{1}{2} \alpha e_{ijk} F^{jk}.
\end{equation}
Here, $e_{ijk}$ is the completely anti-symmetric Levi-Civita pseudo-tensor of the absolute space such that $e_{r\theta\varphi} = \sqrt\gamma$, where, $\gamma$ is the determinant of $ds_{sp}^2$. In addition, we define the shift 3-vector as
\begin{equation}
\beta = (0,\; 0,\; \beta_\varphi)\;.
\end{equation}
The constitutive equations relating the electromagnetic field
vectors, $E$ and $B$, and their duals, $D$ and $H$, for regions of
zero electric and magnetic susceptibilities are
\begin{equation}
E = \alpha D + \beta \times B\;, \;{\rm and} \;H = \alpha B - \beta \times D\;,
\label{suscept}
\end{equation}
where the curl is defined by the equation
$$(A \times B)^i = e^{ijk} A_j B_k\;.$$
With these definitions, Maxwell's equations become
\begin{equation}
\bar \nabla \cdot B = 0  \;, \;{\rm and} \;   \partial_t B +  \bar \nabla \times E = 0
\end{equation}
for the homogenous part and
\begin{equation}
\bar \nabla \cdot D = \rho_c \;, \;{\rm and} \; -\partial_t D +  \bar \nabla \times H = J
\label{max3and4}
\end{equation}
for the inhomogeneous part. Here,
\begin{equation}
\rho_c = \alpha I^t\;,\;\;\;J^k = \alpha I^k \;,
\label{rhoji}
\end{equation}
are the charge and current densities respectively, $\bar \nabla$
is the covariant derivative induced by the spatial metric $\gamma_{ij}$
on our absolute space, and $\alpha$ is the function that appears in
eq.\ (\ref{metfcnalpha}).

\vskip0.25in
\section{Stationary, Axisymmetric, Force-Free Electrodynamics}
\label{fofee}
In the Blanford-Znajek mechanism, the magnetosphere is assumed to be force-free. 
In addition, Blandford and Znajek impose, for simplicity, the inherent symmetry of 
the geometry of the rotating black hole onto the currents and fields, 
namely, stationarity and axisymmetry.
Consequently the Lorentz force on
the currents in the magnetosphere is trivial, so that
\begin{equation}
F_{\mu\nu}I^\nu =0\;.
\label{fofreecovariant}
\end{equation}
In our $3+1$ language, the above condition takes the form
\begin{equation}
E \cdot J=0\;,
\label{fofreecons1}
\end{equation}
and
\begin{equation}
\rho_c\; E + J \times B=0.
\label{fofreecons2}
\end{equation}
The consequences of eqs.\ (\ref{fofreecons1}) and (\ref{fofreecons2})
along with the requirements of stationarity and axisymmetry will be
summarized below. The interested reader is referred to \cite{md05}
for further details. The poloidal and toroidal components ($A_P$
and $A_T$, respectively) of a vector field are defined such that $A = A_P + A_T$,
where $A_P = A^r \partial_r + A^\theta \partial_\theta$ and $A_T =
A^\varphi \partial_\varphi$. Since the magnetosphere is stationary
and axisymmetric,
\begin{equation}
E_T = 0\;.
 \label{etor}
\end{equation}
This is true because stationarity implies that $E$ is the gradient of a scalar,
and for axisymmetric fields, all $\varphi$ derivatives vanish.
From eq.\ (\ref{fofreecons2}) we have that
\begin{equation}
E \cdot B = 0.
\label{transverse}
\end{equation}
Eqs.\ (\ref{etor}) and (\ref{transverse}) imply
that there exists a vector $\omega = \Omega \;\partial_\varphi$ such that
\begin{equation}
E = -\, \omega \times B.
\label{edef}
\end{equation}
Here, $\Omega$ is a function dependent on $r$ and $\theta$. 
Since $B$ is divergence free,
\begin{equation}
B = \bar \nabla \times A\;,
\end{equation}
where $A$ is the 3-vector potential associated with $B$. Therefore,
\begin{equation}
B_P = \frac{1}{\sqrt\gamma}(A_{\varphi,\theta}  \partial_r  -A_{\varphi,r}
 \partial_\theta)\;,
\label{bpexplicit}
\end{equation}
since $A_\varphi$ is independent of $t$ and $\varphi$. Surfaces of constant $A_\varphi$ are called {\it equipotential surfaces}. Here $A_\varphi$ is the toroidal component of the 3-vector potential $A$. Functions that are constant on equipotential surfaces are called {\it equipotential surface functions}, i.e., a function $f$ is an equipotential surface function if
\begin{equation}
B_P \;f = 0 = A_{\varphi,\theta} \;\partial_r f - A_{\varphi,r}\; \partial_\theta f\;.
\end{equation}
It can be shown that $\Omega$  and $H_\varphi$ are equipotential surface functions, and consequently from eq.\ (\ref{edef}) we get
\begin{equation}
E_P = - d\int \Omega \;d A_\varphi\;.
\label{eformula}
\end{equation}
Since $\Omega$ is an equipotential surface function, it is strictly a function of $A_\varphi$, and hence the above integral is well defined. The constitutive eqs.\ (\ref{suscept}) give
\begin{equation}
D= D_P = -\frac{1}{\alpha}(\Omega + \beta^\varphi) \; d A_\varphi\;,
\label{dformula}
\end{equation}
and 
\begin{equation}
H_P = (\alpha^2-\beta^2-\beta_\varphi \Omega)\frac{B_P}{\alpha}\;.
\label{hpolformula}
\end{equation}
The electric charge density is determined by the divergence of
$D_P$:
\begin{equation}
-\sqrt \gamma \rho_c = \partial_r[\frac{1}{\alpha \sqrt{\gamma}}
(\gamma_{\varphi \varphi} \Omega +\beta_\varphi) \;\gamma_{\theta
\theta}\; A_{\varphi,r}]\;+\partial_\theta [\frac{1}{\alpha
\sqrt{\gamma}} (\gamma_{\varphi \varphi}\Omega
+\beta_\varphi)\;\gamma_{r r} \;A_{\varphi,\theta}].
\label{rhoformula}
\end{equation}
The toroidal current vector can be obtained by computing the
$\varphi$ component of the curl in eq.\ (\ref{max3and4}):
\begin{equation}
-\sqrt \gamma J^\varphi = H_{r,\theta} - H_{\theta,r} =
\partial_r[\frac{1}{\alpha \sqrt{\gamma}}
(\alpha^2-\beta^2-\beta_\varphi \Omega) \gamma_{\theta \theta}
A_{\varphi,r}]+\;\partial_\theta [\frac{1}{\alpha \sqrt{\gamma}}
(\alpha^2-\beta^2-\beta_\varphi \Omega)\gamma_{r r}
A_{\varphi,\theta}]. \label{jphiformula}
\end{equation}
The poloidal current is completely determined by $H_\varphi$ and is given by
\begin{equation}
\sqrt \gamma J_p = H_{\varphi,\theta} \partial_r -H_{\varphi,r} \partial_\theta.
\label{jpolformula}
\end{equation}
Finally,
$$B_\varphi = H_\varphi /\alpha.$$
Since $H_\varphi$ is an equipotential surface function, from eq.\ (\ref{jpolformula}) we have that
\begin{equation}
J_P=\frac{1}{\sqrt \gamma} \frac{d H_\varphi}{d A_\varphi}(A_{\varphi,\theta} \partial_r-
A_{\varphi,r} \partial_\theta)=\frac{d H_\varphi}{d A_\varphi}B_P\;,
\label{jpbexplicit}
\end{equation}
i.e., the currents flow along equipotential surfaces.
The only remaining requirement in the explicit formulae given above for the fields and current 
stems from eq.\ (\ref{fofreecons2}), and can be written in the form
\begin{equation}
\frac{1}{2} \frac{d H_\varphi^2}{d A_\varphi }=-\alpha (\rho_c\; \Omega \;\gamma_{\varphi \varphi}-J_\varphi)\;.
\label{finalconsb}
\end{equation}
Since the left hand side of the equation above is an equipotential surface function, we must pick an equipotential surface function $\Omega$ such that the right hand side of the equation is an equipotential surface function as well. Additionally, for the fields and currents to remain well defined at the event horizon of the Kerr black hole, given by
$$r = r_+ \equiv M + \sqrt{M^2-a^2}\;,$$
we must have that
\begin{equation}
H_\varphi \mid_{r_+} = \frac{\sin^2\theta}{\alpha} B^r (2Mr \;\Omega -a)\mid_{r_+}
\label{znaregcond}
\end{equation}
in Boyer-Lindquist coordinates. This is the Znajek regularity
condition derived in \cite{zna77}. In the above equation, all
quantities are to be evaluated at the event horizon.
\vskip0.25in The
poloidal component of the Poynting vector for the electromagnetic
field is given by
$$S_P = (E\times H)_P\;.$$
Consequently, the rate of extraction of energy from the black hole is given by
\begin{equation}
\frac{d  E_{em}}{dt} =   - \int_{r_2=\;\infty}  \Omega H_\varphi B^r \sqrt{\gamma}\;d\theta \;d\varphi\;.
\label{engextractformula}
\end{equation}
In a similar manner, using the fact that $\partial_\varphi$ is a Killing vector field of the geometry, 
we get the following expression for the rate of extraction of angular momentum from the black hole:
\begin{equation}
\frac{d L_{em}}{dt} =    - \int_{r_2=\;\infty}   H_\varphi B^r  \;\sqrt{\gamma}\;d\theta \;d\varphi\;.
\label{angextractformula}
\end{equation}
Under this circumstance, the mass and angular momentum of the black hole decreases at a rate given by
\begin{equation}
\frac{\delta M}{\delta t}=- \frac{d E_{em}}{d t}, {\rm~and~}
\frac{\delta J}{\delta t}=- \frac{d L_{em}}{d t}.
\label{bholerates}
\end {equation}

\section{The general form of Geodesic Currents}
\label{geocurrent}
In the case of force-free electrodynamics in a curved spacetime, 
we would expect the charged particles to flow along the geodesics of the geometry 
(a fact that was manifest in eq.\ (\ref{oldcurrent})). With this in mind, we write the current vector in the form
\begin{equation}
I^{\nu} = F(r, \theta)\; u^{\nu}\;.
\label{current}
\end{equation}
Here, $u^{\nu}$ is a timelike geodesic of the Kerr geometry. The function $F$ relates the 4-velocity of the charged particle at a point to electromagnetic current density at that point.  The properties of $u^{\nu}$ are well understood (for example see \cite{CH92} and \cite{ON95}). In particular, $u^{\nu}$ can be written in the form
\begin{equation}
u^{\nu} = (\dot{t}, \frac{\sqrt{R}}{\rho^2},\frac{\sqrt{\Theta}}{\rho^2},\dot{\varphi})\;.
\label{geodesic}
\end{equation}
Here, $\dot{t}$, $\dot{\varphi}$, $\sqrt{R}$ and $\sqrt{\Theta}$ are functions of $r$ and $\theta$. Also, in $\sqrt{R}$ and $\sqrt{\Theta}$
 we haven't yet committed to the ``+" root.
 The notion of time (dot) derivative is not relevant to our analysis;
 the notation has been simply borrowed from the existing literature on Kerr geodesics. The explicit forms of the geodesic functions are
\begin{equation}
\dot t = \frac{\Sigma^2 {\cal E} - 2 a M r {\cal L}}{\rho^2 \Delta},
\label{tdot1}
\end{equation}
and
\begin{equation}
\dot \varphi = \frac{2 a M r {\cal E} \sin^2 \theta + (\rho^2 - 2 M r ){\cal L}}{\rho^2 \Delta \sin^2 \theta}\;.
\label{phidot1}
\end{equation}
The functions $R$ and $\Theta$ are given by
\begin{equation}
R= C^2 + \Delta (q^2 r^2 - {\cal K}) \label{rfunct}
\end{equation}
and
\begin{equation}
\Theta =  {\cal K} +q^2a^2 \cos^2 \theta -\frac{D^2}{\sin^2
\theta}\;, \label{thetafunct}
\end{equation}
where
\begin{equation}
C = (r^2+a^2){\cal E} -a {\cal L} \; \;\;{\rm and }\;\;\; D  = {\cal L} - a {\cal E} \sin^2\theta\;.
\label{geoterms1}
\end{equation}
In the study of Kerr geodesics, ${\cal E}$, ${\cal L}$, and $\cal K$ are simply constants for a given geodesic, and
$$q^2 = 1,\; 0,\; -1$$
for spacelike, null, and timelike geodesics respectively. The terms ${\cal E}$ and ${\cal L}$ can be related to the energy and angular momentum of the particle that flows along the geodesic. Explicitly,
\begin{equation}
{\cal E} = - g(K, u) = - (g_{tt} \dot t + g_{t \varphi} \dot \varphi) \;
\label{engdef}
\end{equation}
and
\begin{equation}
{\cal L} =  g(m , u) = (g_{t \varphi} \dot t + \gamma_{\varphi \varphi} \dot \varphi)\;.
\label{angdef}
\end{equation}
Here $K = \partial_t$ and $m=\partial_\varphi$.
${\cal K}$ is the famous Carter's constant \cite{Cart68}. In our case, since the currents 
flow along equipotential surfaces, ${\cal E}$, ${\cal L}$ and ${\cal K}$ become equipotential surface functions, which are now subject to the condition
\begin{equation}
\sqrt{R}\; f_{,r}\;+\;\sqrt{\Theta}\;f_{,\theta} = 0\;,
\end{equation}
for any equipotential surface function $f$. Also, since we are only interested in particles with mass, we will set $q^2 = -1$.
Conservation of electric charge is implied by the equation $\nabla_{\mu} I^{\mu}=0$, i.e.,
\begin{equation}
\partial_r[ F \frac{\sqrt{R}}{\rho^2}\sqrt{-g}]+\partial_\theta[F \frac{\sqrt{\Theta}}{\rho^2}\sqrt{-g}]=0\;.
\label{conserve}
\end{equation}
This is actually an equation for $F$.

In writing eq.\ (\ref{current}), we have assumed the explicit form of the charge density and the 3-vector current density. In the remainder of this section, we will construct a formula for $H_\varphi$ and show that it satisfies the Znajek regularity condition, eq.\ (\ref{znaregcond}). But first we present a few preliminaries.

From eq.\ (\ref{rhoji}) we see that
$$\rho_c = \alpha F \;\dot{t}\;,\;\;\;\;J^\varphi = \alpha F \;\dot{\varphi}\;,$$
and
\begin{equation}
J_P = \alpha F \;\frac{\sqrt{R}}{\rho^2}\;\partial_r  \; + \; \alpha F \; \frac{\sqrt{\Theta}}{\rho^2} \; \partial_\theta\;.
\label{geochargeandcurrents}
\end{equation}
Consequently, eq.\ (\ref{finalconsb}) can be written as
\begin{equation}
-\frac{1}{2} \frac{d H_\varphi^2}{d A_\varphi}= \Delta \sin^2\theta F ( \dot{t}\; \Omega - \dot{\varphi}).
\label{constgeocurrent}
\end{equation}
In the theorem that follows, we obtain an explicit expression for
$H_\varphi$.
\vskip0.2in
\begin{thm}
\label{hphiformula} {\rm In force-free, stationary,
axisymmetric electrodynamics in a Kerr background when the
electromagnetic current vector takes the form given by
eq.\ (\ref{current})},
\begin{equation}
H_\varphi = \frac{\Delta \sin\theta \;(\dot t\;\Omega - \dot \varphi)}{\sqrt\Theta}\;A_{\varphi,r}\;.
\label{htorform}
\end{equation}
\end{thm}
Proof: From eqs.\ (\ref{jpolformula}) and (\ref{geochargeandcurrents}) we get that
\begin{equation}
H_{\varphi,r} = - F \sin\theta \sqrt \Theta\;.
\label{hphir}
\end{equation}
But note also that
$$-H_\varphi \; H_{\varphi,r} = -\frac{1}{2} \frac{d H_\varphi^2}{d A_\varphi} A_{\varphi,r}\;.$$
Substituting eqs.\ (\ref{constgeocurrent}) and (\ref{hphir}) into the
above equation, we get the necessary result.
\qed
\vskip0.15in

Since
$A_\varphi$ is an equipotential surface function, $H_\varphi$ can
also be written in the form
\begin{equation}
H_\varphi = \frac{- \; \Delta \sin\theta \;(\dot t\;\Omega - \dot \varphi)}{\sqrt R}\;A_{\varphi,\theta}\;.
\label{htoralternate}
\end{equation}

\vskip0.15in
\begin{cor}
{\rm $H_\varphi$ as given by {\rm eq.\ (\ref{htorform})} 
and/or {\rm eq.\ (\ref{htoralternate})} will be an equipotential surface function if and only if
\begin{equation}
R \;\partial_r\;(\frac{\chi}{\sqrt R})\;+\;\Theta \;\partial_\theta\;(\frac{\chi}{\sqrt \Theta} )= 0\;,
\label{hphiequipot}
\end{equation}
where
$$\chi = \Delta \sin\theta \;(\dot t\;\Omega - \dot \varphi)\;.$$}
\end{cor}
Proof:
From eqs.\ (\ref{htorform}) and (\ref{htoralternate}),
$$H_\varphi = \frac{\chi}{\sqrt \Theta}\;A_{\varphi,r} =  -\frac{\chi}{\sqrt R}\;A_{\varphi,\theta}\;,$$
and since $H_\varphi$ is to be an equipotential surface function
$$\sqrt R\;H_{\varphi,r}\;+\;\sqrt \Theta\;H_{\varphi,\theta}=0\;.$$
Since $A_\varphi$ is an equipotential function, the above two equations immediately imply eq.\ (\ref{hphiequipot}).
\qed
\vskip0.15in One of the main advantages, and justification for
eq.\ (\ref{current}) is that the expression for $H_\varphi$ as derived
above naturally satisfies the Znajek regularity condition.
\vskip0.15in
\begin{thm}
{\rm The form of $H_\varphi$ as given in {\rm
eq.\ (\ref{htoralternate})} (or equivalently in {\rm
eq.\ (\ref{htorform})}) satisfies the Znajek regularity condition, 
eq.\ (\ref{znaregcond}), when
$$\sqrt{R}|_{r_+} = -C\;.$$
Note, the only choices for $\sqrt{R}|_{r_+}$ are $\pm\; C$.}
\label{znacondforfree}
\end{thm}
Proof:
From eq.\ (\ref{bpexplicit}), $$\frac{A_{\varphi,\theta}}{\sqrt \gamma } = B^r\;.$$
Also,
$$H_\varphi = \frac{- \; \Delta \sin\theta \;(\dot t\;\Omega - \dot \varphi)}{\sqrt R}
\;A_{\varphi,\theta}$$
$$=\frac{- \sin^2\theta}{\alpha \sqrt R} \;\rho^2 \Delta \;(\dot t\;\Omega - \dot \varphi) B^r\;.$$
We choose the square root such that
$$\sqrt R |_{\Delta = 0} = -\;C|_{\Delta = 0}\;.$$
Here, $C$ is the function defined in eq.\ (\ref{geoterms1}).
Also,
$$\rho^2 \Delta\; \dot t |_{\Delta = 0} = (r_+ ^2 + a^2)^2 {\cal E} - 2 a M r_+ \;{\cal L} = 2  M r_+\;C|_{\Delta = 0}\;,$$
and
$$\rho^2 \Delta\; \dot \varphi |_{\Delta = 0} = a \;C|_{\Delta = 0}\;.$$
Putting all the above equations together, we get the stated result.
\qed
\section{The Extraction of Energy and Angular Momentum via Particle Jets}
\label{particlejets}

In Section \ref{fofee}, the terms representing the extraction of
energy and angular momentum (eqs.\ (\ref{engextractformula}) and
(\ref{angextractformula})) were derived from the conserved
quantities of the electromagnetic stress tensor. In this section, we
will do the same for the matter fields. This will be useful for
calculating the energy released via matter jets from supermassive
black holes. Separate conservation of field and particle energy and
angular momentum stems from the fact that we are only considering a
force-free magnetosphere. In particular, the total energy-momentum
tensor of magnetohydrodynamics can be written in the form:
$$T^{\mu\nu}= T^{\mu\nu}_{em}\;+\;T^{\mu\nu}_{matter}\;.$$
Since this must be divergence free, we must have that
$$\nabla_\mu T^{\mu\nu}= \nabla_\mu T^{\mu\nu}_{em}\;+\; \nabla_\mu \;T^{\mu\nu}_{matter} = 0\;.$$
But, from eq.\ (\ref{fofreecovariant}) we see that
$$\nabla_\mu T^{\mu\nu}_{em} = 0\;.$$
Consequently, we must have that
$$\nabla_\mu T^{\mu\nu}_{matter} = 0\;,$$
as mentioned above. When the magnetosphere is filled with a geodesic congruence, we can take the matter energy momentum tensor as
$$ T^{\mu\nu}_{matter} = (m/e)\; F (r,\theta)\; u^\mu\; u^\nu\;,$$
where $e$  and $m$ are the electric charge  and mass of the particle species, respectively, 
and $F$ and $u^\mu$ are as given in eq.\ (\ref{current}). For then,
$$\nabla_\mu T^{\mu\nu}_{matter} =$$
$$ (m/e) \;\nabla_\mu (F \; u^\mu) u^\nu\ + (m/e) \;F\; u^\mu \nabla_\mu  u^\nu\;=\;0\;.$$
In the equation above, the first term on the right hand side
vanishes due to the continuity equation for the electromagnetic
current density, and the second term is trivial because of the
geodesic equation. If the charge of the currents are position
dependent, we must either take the derivative of the $e/m$ term in
the above equation, or consider the regions of different values for
$e/m$ separately.
\subsection{Matter Energy Flux}
Since $K$ is a Killing vector field of the Kerr geometry,
$$\nabla_\mu \;(T^\mu_{\nu\; matter} \; K^\nu) = 0\;,$$
i.e.,
$$\frac{1}{\alpha\;\sqrt{\gamma}} \partial_t (\alpha\;\sqrt{\gamma}\; T^t _{t\; matter})\;+\;\frac{1}{\alpha\;\sqrt{\gamma}}
\partial_i (\alpha\;\sqrt{\gamma} \;T^i _{t\; matter}) = 0\;.$$
Let
\begin{equation}
e = - \;\alpha \; T^t _{t\;matter}
\end{equation}
be the energy density of the matter fields, and
\begin{equation}
S^i  = - \;\alpha \; T^i _{t\;matter}
\end{equation}
be the matter energy density flux. Therefore,
\begin{equation}
\frac{d}{dt}\int_V e \;d V = - \int_V \nabla \cdot S \;dV = - \int_{\partial V} S \cdot n \;dA\;.
\end{equation}
Here, $V$ is a 3-dimensional manifold in our absolute space, $\partial V$ is the boundary of $V$, and $n$ is the unit outward pointing normal on $\partial V$ as per the divergence theorem.
Consequently, the rate of matter extraction of energy from the black hole is given by
\begin{equation}
\frac{d  E_m}{dt} =  \int_{r_2=\;\infty} S^r \; \sqrt{\gamma}\;d\theta \;d\varphi\;.
\end{equation}
Clearly,
\begin{equation}
 S^r  = -\;\alpha \;\frac{m}{e}\;F \; \frac{\sqrt R}{\rho^2} (g_{tt} \;\dot t + g_{t\varphi}\;\dot\varphi) \;.
\end{equation}
But
$$- (g_{tt} \;\dot t + g_{t\varphi}\;\dot\varphi)$$
is nothing more than the poloidal function ${\cal E}$, therefore,
\begin{equation}
\frac{d  E_m}{dt} =  \frac{m}{e} \int_{r_2=\;\infty} F \; \sqrt R {\cal E}\; \sin\theta \;d\theta \;d\varphi\;.
\label{matterengextractformula}
\end{equation}
\subsection{Matter Angular Momentum Flux}
Using the Killing vector field $m \equiv \partial_\varphi$, we can deduce a similar expression for the rate of extraction of angular momentum from the black hole via particle jets.
Setting
\begin{equation}
l =  \alpha \; T^t _{\varphi \;matter}
\end{equation}
as the angular momentum density of the matter fields, and
\begin{equation}
L^i  =  \alpha \; T^i _{\varphi \;matter}
\end{equation}
as the matter angular momentum density flux, we see that the rate of extraction of angular momentum from the black hole via particle jets is given by
\begin{equation}
\frac{d  L_m}{dt} =  \frac{m}{e} \int_{r_2=\;\infty} F \; \sqrt R {\cal L}\; \sin\theta \;d\theta \;d\varphi\;.
\label{matterangformula}
\end{equation}
\section{The Area Theorem, and the Blandford-Znajek Mechanism}
\label{sectarea}

\noindent In \cite{bz77}, it was shown that for electromagnetic
extraction of energy from the horizon of the black hole, the rate of
energy extraction was related to the rate of angular momentum
extraction by the expression
\begin{equation}
\frac{d  E_{em}}{dt} \leq  \Omega_H\;\frac{d  L_{em}}{dt}\;.
\end{equation}
Here,
\begin{equation}
\Omega_H = \;\frac{a}{r_+ ^2 + a^2}
\end{equation}
is the angular velocity of the event horizon. A black hole that is adiabatically evolving along a Kerr sequence
such that its rate of energy and angular momentum extraction satisfies
\begin{equation}
\frac{d  E}{dt} \leq  \Omega_H\;\frac{d  L}{dt}
\label{extractrate}
\end{equation}
cannot lead to the formation of a naked singularity \cite{chr70}. This can be easily seen as follows.
The irreducible mass $M_{irr}$ of a Kerr Black hole is given by
\begin{equation}
M_{irr}^2 = \frac{1}{2}[M^2 + \sqrt{M^4 - J^2}]\;,
\end{equation}
where $J = a M$ is the angular momentum of the black hole. Under extraction rates subject to eq.\ (\ref{extractrate}), the mass and the angular momentum of the Black hole decreases to
\begin{equation}
M \rightarrow M + \delta M\;\;\;{\rm and}\;\;\;J \rightarrow J + \delta J\;,
\end{equation}
where $ \delta M$ and $ \delta J$ satisfy
\begin{equation}
 \delta M \geq  \Omega_H\; \delta J\;.
\label{bhextractrate}
\end{equation}
We now state the following calculation due to Christodoulou  \cite{chr70}) without proof.
\begin{thm}
{\rm When energy and angular momentum is extracted from a Kerr black hole so that
eq.\ (\ref{bhextractrate}) (or equivalently eq.\ (\ref{extractrate})) is satisfied, we have that} $\delta M_{irr}^2 \geq 0\;.$
\end{thm}
The area of the event horizon is given by
\begin{equation}
A = \int_{r_+}\sqrt{g_{\theta \theta}g_{\varphi \varphi}}\; d\theta d \varphi = 16 \pi M_{irr}^2\;.
\label{aeh}
\end{equation}
Since the area of the event horizon is proportional to $M_{irr}^2$,
from the theorem above, we see that it is non-decreasing,  and so,
we have the following corollary.
\begin{cor}
{\rm A black hole that is adiabatically evolving along a Kerr
sequence such that its rate of energy and angular momentum
extraction satisfies eq.\ (\ref{extractrate}) always contains an event
horizon that cloaks the physical singularity of the Kerr geometry.}
\label{nakedsingcor}
\end{cor}

Of course, when all the angular momentum is extracted, the Kerr
black hole reduces to the Schwarzschild solution, and here too we
do not have a naked singularity. A dynamically evolving black hole, in general, need not follow a Kerr sequence.
In this case, the powerful area theorem of Stephan Hawking \cite{Haw71} requires that the area of the event horizon continues to be non-decreasing.

We conclude our analysis of the geodesics currents in the force-free magnetosphere of the Kerr geometry by showing that the matter energy and angular momentum extraction rates are subject to the same inequality as their electromagnetic counterparts (eq. (\ref{extractrate})), and hence we are in no danger of violating the Penrose conjecture.

\noindent \begin{thm}
{\rm When the currents in a force-free magnetosphere are given by   eq.\ (\ref{current}), the extraction of energy from the matter fields will not lead to the formation of a naked singularity.}
\end{thm}
Proof:
At the horizon, if charge are to flow out to the magnetosphere, from theorem (\ref{znacondforfree}) we need that
$$\sqrt{R}|_{r_+} = -C > 0\;.$$
This is possible only when the geodesics are such that
\begin{equation}
{\cal E} \leq  \Omega_H\;{\cal L}\;,
\end{equation}
and consequently from eqs.\ (\ref{matterengextractformula}) and (\ref{matterangformula})
\begin{equation}
\frac{d  E_{m}}{dt} \leq  \Omega_H\;\frac{d  L_{m}}{dt}\;.
\end{equation}
Therefore, corollary (\ref{nakedsingcor}) implies the necessary result.
\qed

\section{Conclusions}
\label{conclusions}

The mechanism to form jets from black holes is a major unsolved problems in astrophysics, 
and its explanation is important for our understanding of nonthermal 
radiation form Solar-mass black holes 
in the Milky Way, newly born black holes associated with gamma-ray bursts, 
and supermassive black holes at the centers of distant galaxies. 
One possibility is that black holes with radio-emitting 
jets are distinguished from radio-quiet black holes by extracting 
rotational energy from the black hole through the Blandford-Znajek process. 
Matter currents can also carry energy and angular momentum, and represent a plausible 
alternative to electromagnetic Poynting flux for extracting the black hole's rotational energy.

In this paper, we have used the $3+1$ formulation of general relativity
to analyze currents that flow in the force-free
magnetosphere near a Kerr black hole.  We have rewritten 
the constraint equation for a force-free magnetosphere, eq.\ (\ref{finalconsb}), 
to apply to particles that follow timelike geodesics, yielding 
eq.\ (\ref{constgeocurrent}).
The toroidal magnetic susceptibility, $H_\varphi$, from which the poloidal current
is derived, is shown to satisfy the Znajek regularity condition. Expressions 
for energy and angular momentum flux associated with the matter currents 
were obtained. We show that extraction of energy and angular momentum cannot lead to the 
formation of a naked singularity. In future work, trajectories of particles in 
the black hole magnetosphere will be derived, and will be compared with 
observations of jets from black holes.

\vskip0.5in
G.\ M.\ acknowledges funding through a Troy University faculty development grant.
This research is also supported
through NASA {\it GLAST} Science Investigation No.\ DPR-S-1563-Y.
The work of C.\ D.\ D.\ is supported by the Office of Naval Research.

\end{document}